\documentclass[conference]{IEEEtran}
\IEEEoverridecommandlockouts
\usepackage{cite}
\usepackage{amsmath,amssymb,amsfonts}
\newtheorem{definition}{Definition}
\usepackage{algorithmic}
\usepackage{graphicx}
\usepackage{textcomp}
\usepackage{xcolor}
\def\BibTeX{{\rm B\kern-.05em{\sc i\kern-.025em b}\kern-.08em
    T\kern-.1667em\lower.7ex\hbox{E}\kern-.125emX}}
\begin{document}

\title{Active Region-based Flare Forecasting with Sliding Window Multivariate Time Series Forest Classifiers}

\author{\IEEEauthorblockN{Anli Ji and Berkay Aydin}
\IEEEauthorblockA{\textit{Department of Computer Science} \\
\textit{Georgia State University}\\
Atlanta, GA \\
\{aji1, baydin2\}@gsu.edu}

}

\maketitle

\begin{abstract}
Over the past few decades, many applications of physics-based simulations and data-driven techniques (including machine learning and deep learning) have emerged to analyze and predict solar flares. These approaches are pivotal in understanding the dynamics of solar flares, primarily aiming to forecast these events and minimize potential risks they may pose to Earth. Although current methods have made significant progress, there are still limitations to these data-driven approaches. One prominent drawback is the lack of consideration for the temporal evolution characteristics in the active regions from which these flares originate. This oversight hinders the ability of these methods to grasp the relationships between high-dimensional active region features, thereby limiting their usability in operations. This study centers on the development of interpretable classifiers for multivariate time series and the demonstration of a novel feature ranking method with sliding window-based sub-interval ranking. The primary contribution of our work is to bridge the gap between complex, less understandable black-box models used for high-dimensional data and the exploration of relevant sub-intervals from multivariate time series, specifically in the context of solar flare forecasting. Our findings demonstrate that our sliding-window time series forest classifier performs effectively in solar flare prediction (with a True Skill Statistic of over 85\%) while also pinpointing the most crucial features and sub-intervals for a given learning task.
\end{abstract}


\section{Introduction}
\begin{figure}[b]
    \centering
    \includegraphics[width=.485\textwidth]{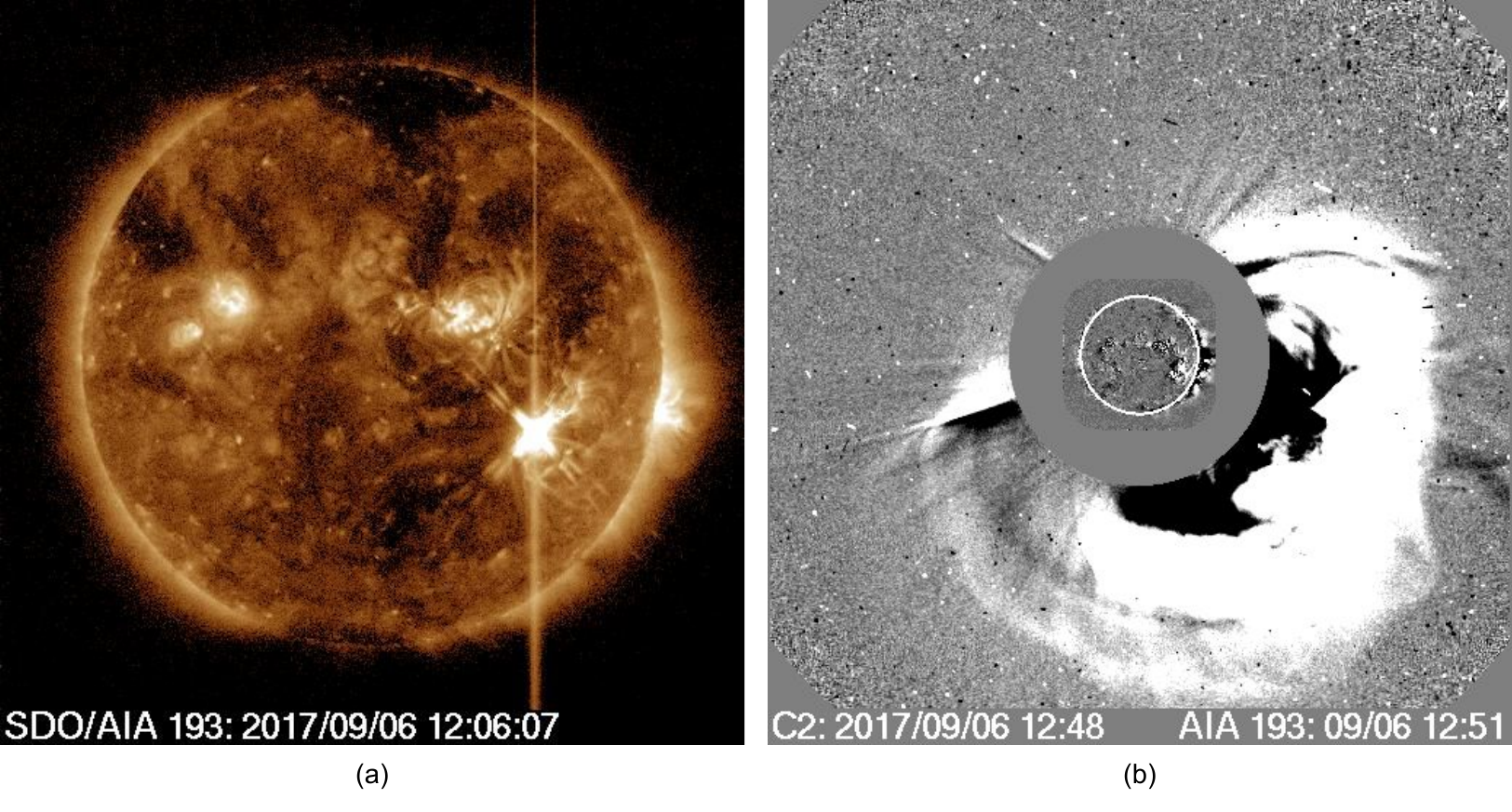}
    \caption{An X9.3-class flare occurred in 2017-09-06 captured by Solar Dynamics Observatory (a) and its accompanying coronal mass ejection captured by Large Angle and Spectrometric Coronagraph (LASCO) instrument (b)}
    \label{fig:flcme}
\end{figure}

Solar flares are intense localized bursts of electromagnetic energy originating from the Sun's atmosphere, varying in duration from minutes to hours. These sudden energy releases travel nearly at the speed of light and can coincide with other eruptive solar events such as coronal mass ejections (CMEs). These events have the potential to trigger severe geomagnetic storms, cause extensive radio disruptions on the sunlit side of Earth, and disturb sensitive space equipment close to Earth. An illustrative example of an X9.3-class solar flare (on September 6, 2017) accompanied by a CME is shown in Figure~\ref{fig:flcme}, which led to an Earth-impacting solar energetic particle event the same day. Since solar flares are central events, efforts to forecast them, combining physics-based and data-driven methodologies, have been prominent in the literature \cite{priest2002magnetic} \cite{shibata2011solar} \cite{Kusano2020}. These methods often leverage data from solar magnetograms gathered by diverse satellites and employ diverse learning algorithms such as support vector machines (SVMs) \cite{Bobra2016} \cite{Boucheron2015}, logistic regression \cite{Song2008}, and decision trees \cite{Yu2009} \cite{Huang2010}.

Recently, many methodologies and corresponding models have been introduced to tackle flare prediction as a classification task by using point-in-time measurements. However, these approaches usually overlook the inherent time-dependent (i.e., temporal evolution characteristics) nature of the data (see \cite{Georgoulis2012} for a review), treating multiple physical observations as separate entities and generating predictions solely based on instantaneous values \cite{Boucheron2015}. This disregards the dynamic essence of flares, whose characteristics depend on the evolutionary behavior of solar active regions \cite{Benz2008} \cite{Fletcher2011}. By exploring temporal features spread across time series intervals, we can potentially unveil implicit relationships and capture relevant patterns underneath. In our earlier work \cite{9377906}, we employed interval-based time series classification models on multivariate time series data using a set of interval features extracted. However, the computational efficiency of this approach led to the extraction of interval features from randomly sampled interval sets, lacking a systematic evaluation of the internal process of determining features (or intervals from these time series features) that are more important for predictive tasks. Hence, our primary focus in this study is interpreting statistical features obtained from multivariate time series and identifying sequences of features (and related patterns) using multi-scale sliding windows with varying interval windows and step sizes.

The rest of the paper is structured as follows: Section \ref{related_work} provides background information on existing time series classification models pertinent to flare prediction. In Section \ref{methodology}, we provide our problem formulation and introduce our multivariate time series classification model and feature ranking method used for extracting relevant feature intervals from provided time series data. Our experimental setup and evaluation framework are presented in Section \ref{experiments}. Finally, Section \ref{conclusions} provides conclusions drawn from our study and discusses potential avenues for future research.

\section{Related Work}\label{related_work}
Within the proliferation of available time series data sets \cite{Silva2018} and a wide spectrum of machine learning-based techniques proposed for time series classification, two notable categories utilized for these predictive tasks are similarity-based and feature-based algorithms. The former group makes predictions by measuring the similarity between training and testing instances, while the latter generates predictions by utilizing temporal features extracted from entire time series or subsequences within these time series instances. Especially in the task of solar flare prediction or other tasks (such as anomaly detection \cite{Homayouni2020}), approaches proposed utilizing either full-disk (e.g., \cite{9671322}\cite{Pandey2022} \cite{pandey2023interpretable}) or active region-based (e.g., \cite{9377906} \cite{9750381} \cite{9378006} \cite{Hong2023}) patches have shown great impact with respect to derived time series features (i.e., features extracted from each subsequent time series). Among the commonly employed similarity-based algorithms, some rely on nearest-neighbor classifiers combined with metrics like Euclidean distance (NN-Euclidean) or elastic measures like Dynamic Time Warping (NN-DTW) \cite{6497440}, \cite{1163055}, \cite{Bagnall2016},\cite{Lines2014}. In contrast, feature-based algorithms employ derived time series features to capture associations between target variables and time series instances. For instance, in \cite{10.5555/766914.766918}, basic statistical features (e.g., mean and standard deviation) were extracted from global time series and used as input for a multi-layer perceptron network, yet neglecting localized informative characteristics and properties. Meanwhile, \cite{Geurts2001} considers local attributes through piecewise constant modeling, coupled with extracted patterns to enhance model interpretability. However, this approach also comes with the issue of acquiring simplistic features during the feature selection process. In \cite{DBLP:conf/pkdd/EruhimovMT07}, an extensive number of features derived from time series data was incorporated (e.g., wavelets, Chebyshev coefficients, etc.), but the method incurs high computational costs and lacks inherent interpretability in high-dimensional data spaces. 

Feature-based classification methods face challenges when dealing with multivariate time series data, as they require additional intricate information across features. Generating discriminating features in high-dimensional spaces becomes difficult due to the unknown interrelations among input parameters of the multivariate time series, therefore, adding complexity to model construction. Various techniques have been attempted to address this challenge, often utilizing ensembled univariate models across individual feature spaces for predictions. These methods focus on extracting relevant features in multivariate aspects and then applying traditional machine learning algorithms for classification. Common features include statistical measures (e.g., mean, variance), spectral features (e.g., Fast Fourier Transform coefficients), or time-domain features (e.g., autocorrelation). For example, Shapelet-based decision trees \cite{Ye2010} combine the concept of shapelets (i.e., discriminative sub-sequences that capture distinctive patterns in time series data) with the ensemble approach. It extracts shapelets from the training data and constructs an ensemble of decision trees (i.e., random forest) where each estimator focuses on a different subset of shapelets, usually measured by Euclidean distance (i.e., the distance to the finest position in the time series). This method can effectively capture local patterns in multivariate time series data but also can be computationally expensive and hard to identify relevant shapelets (especially in high dimensions) that are both informative and applicable across dimensions. Another problem is that the shapelets extracted from one multivariate time series dataset might not generalize well to other datasets with different dimensionality, characteristics, and patterns. To mitigate these issues, the Generalized random shapelet forest (gRSF) \cite{Karlsson2016} enhances the original shapelet-based method by measuring distances between randomly selected time series and other time series falling within a threshold distance of the representative shapelet. Similarly, the Time Series Forest (TSF) \cite{Deng2013} also incorporates subseries, but instead of relying on distance measurement from learned subsequences, it derives summarized statistical features (such as mean, standard deviation, and gradient) within randomly divided intervals of the univariate time series. It treats each time step as a separate component and constructs decision trees in each feature dimension to capture temporal relationships and reduce the high-dimensional feature space. However, important interrelationships between different components of the time series might not be fully captured by treating them as separate features, which can lead to a loss of inter-channel relationships and dependencies that are often crucial in multivariate time series data. Furthermore, the Canonical Interval Forest (CIF) \cite{DBLP:journals/corr/abs-2008-09172} extends TSF by incorporating additional canonical characteristics of the time series and \emph{catch22} \cite{Lubba2019} features extracted from each interval. This approach is designed to capture both individual patterns within each time series component and relationships between different components but can only provide a certain level of interpretability as interpreting an ensemble of decision trees can be more challenging. Understanding the combined effects of multiple trees on multivariate time series data might be less intuitive compared to single decision trees.


Many of these methods focus solely on understanding how each feature behaves on its own, without considering how different features might interact. There are instances where a particular relationship within a single time series parameter might be significant for a specific feature, but not necessarily for others. The connections between distinct features are often not known upfront. When trying to generate features that effectively differentiate classes, it is more advantageous to select the most relevant time intervals to create a more robust model. However, identifying these relevant intervals is not an easy task as they typically cannot be directly determined and require an expensive search across the entire series. Being able to extract the underlying mutual information present in these relevant intervals can enhance our understanding of the predictive process and accelerate the transition from research to operations in flare forecasting models. Our objective in this work is to establish a framework that can recognize these characteristics and offer deeper insights into the behavior of classifiers during prediction tasks.

\section{Methodology}\label{methodology}
\begin{figure*}
    \centering
    \includegraphics[width=.99\textwidth]{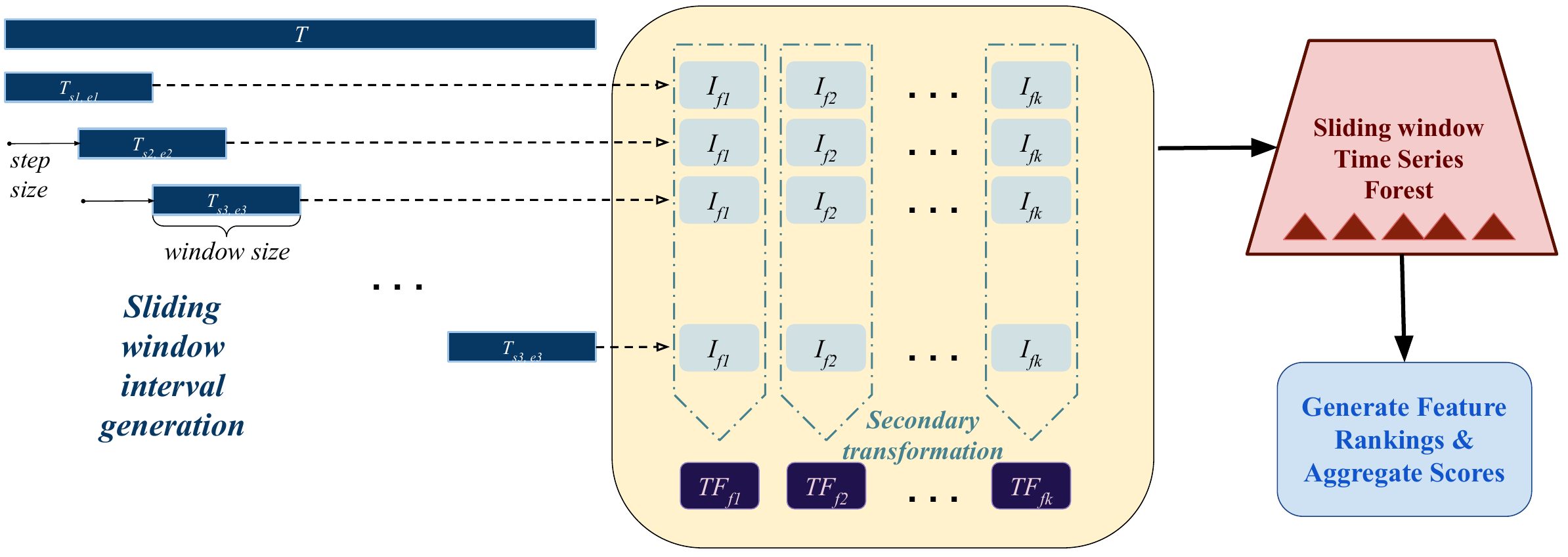}
    \caption{The overview of the methodology presented in this paper. We first generate subsequences (intervals) with a sliding window. Then, we create vectorized features from these intervals where these features can be used as input for the sliding window time series forest (a random forest built on multivariate time series features) and features are ranked with aggregated relevance scores.}
    \label{fig:overview}
\end{figure*}

In this section, we will outline the core aspects of our approach. We will start by defining the problem we are addressing and explaining how we generate statistical features from intervals. Next, we will introduce the sliding-window time series forest, a key component of our method. Lastly, we will describe our feature ranking technique. A visual representation overview of our methodology is provided in Figure ~\ref{fig:overview}.

\subsection{Problem Formulation}
The sliding-window time series forest represents an intricate interval-based classifier for time series, employing a random forest approach on interval features extracted from individual univariate time series. In the following parts, we outline the definitions and explanations concerning the manipulation of multivariate time series data and the construction of a random forest using these transformed features.

\begin{definition}
A time series $T$ is a sequence of real values observed in ascending time order pair such that \(<(t_1, v_1), (t_2, v_2),..., (t_n, v_n) >\) where \(t_1\) to \(t_n\) represent the equally spaced timestamps \((\forall i, t_{i+1} - t_i = \Delta t\) at which the values are being measured. 
\end{definition}

Such values (i.e., \(v_i\)) can be $P$ dimensional at each time point \(t_i\) in terms of multivariate time series (i.e., $mvts$). We refer to each distinctive value in the same dimension as a feature $f$ at the same time point, meaning that each timestamp \(t_i\) has  $P$ features. 

\begin{definition}
A multivariate time series dataset can be defined as \(D_{MVTS} = \{mvts_1, mvts_2,..., mvts_i\}\) where the length of each series \(|mvts_i| = k_i\) such that there are $k$ number of pairs (i.e., \(<(t_1, v_1), (t_2, v_2),..., (t_k, v_k)>\)) included and the value of \(k_i\) can be varied across the dataset.
\end{definition}

To create an equal-length multivariate time series dataset, one approach involves performing a slicing operation by sliding through each \(mvts_i\) with a window size $l$ so that each slice would contain a set of \(<(t_s, v_s),..., (t_{s+l}, v_{s+l})>\) where $s$ refers to the starting time of each slice.

\begin{definition}
An interval \(T_{s,e}\) is defined as a subsequence of the original time series $T$ such that \(<(t_s, v_s), (t_{s+l}, v_{s+l}),..., (t_e, v_e)>\) with starting point $s$ and ending point $e$ where \(1 \leq s < e \leq n\).
\end{definition}

A dataset created using the sliding window technique and intervals can be potentially different, despite both methods involving comparable operations. The former (i.e., by incorporating the sliding window) augments the dataset with more instances while the latter (i.e., by using intervals) encompasses only a segment of the original dataset. It often refers to each instance generated through a sliding window procedure as an interval, marked by its initiation time $s$ and culmination time $s+l$.

For a comprehensive assessment of the impact of a time series feature on a target variable, it is necessary to pinpoint the pertinent intervals and their connections with other time series (and their intervals) within a dataset. Essentially, understanding frequently occurring patterns in a time series classification task allows us to identify strong correlations with others and discover the hidden shared characteristics among them. Nevertheless, these pertinent parameters might encompass only specific intervals rather than the entire time series. In order to carry out classification tasks on a multivariate time series entity $T$, the task of recognizing relevant intervals involves determining a set of ranked intervals based on their significance in contributing to the classification process. To achieve this, it becomes essential to recognize and scrutinize the characteristics of univariate features.

\begin{definition}
A univariate feature interval set (i.e., $UFI$, which we will denote as $I$ for brevity of notation) with a sliced starting point $s$ and ending point $e$ for each univariate feature $f$ is defined as \(I_f = \{T_{s,e}:cv_1, T_{s+1,e+1}:cv_2,..., T_{s+m,e+m}:cv_{m-1}\}\) where $cv$ refers to the contribution value of such interval $T_{s, e}$. For every feature $f_i$ in a multivariate time series dataset $D_{MVTS}$, we can generate a set of $UFIs$ in the form of \(I = \{I_{f_i},..., I_{f_j}\}\) where this set of $UFIs$ can be ordered based on the starting time of each $UFI$ such that \(\{I_{f_i}:s_{f_i}\} < \{I_{f_j}:s_{f_j}\}\).
\end{definition}

In order to assess the relevance of an interval set, it is essential to recognize how the feature impacts the target variable across various parametric settings in the learning process. Several contribution functions exist as alternatives for quantifying this influence, including supervised feature ranking scores, direct rankings, or applying a Boolean function to the top-k ranked list. These individual scores can subsequently be aggregated to determine the overall relevance of each univariate feature interval set ($UFI$) in relation to the classification task.


\subsection{Interval Features}
In order to generate well-ordered and pertinent intervals, we compute interval statistical features (i.e., mean, standard deviation, and slope) along with secondary transformed features (i.e., maximum, minimum, and mean pooled features) for each subsequence. Initially, we obtain sets of potential candidate intervals using pre-defined window and step sizes. Here, the window size corresponds to the total number of equally-spaced timestamps that need to be gathered from the original time series within each slice, while the step size determines the incremental movement of the sliding window from one to the next. Then, by employing a method similar to the one described in \cite{Deng2013}, we compute statistical features (as indicated in Eq.~\ref{eq1:mean}, Eq.~\ref{eq2:std}, and Eq.~\ref{eq3:slope}) from each of these potential candidate interval sets \(\{(t_1, v_1), (t_2, v_2),..., (t_k, v_k)\}\), thereby creating a collection of extracted interval features. Note here that these derived statistical attributes serve as vectorized representations of a subsequence generated through the application of a sliding window mechanism.

\begin{equation}\label{eq1:mean}
f_{mean}(t_1,t_k) = \frac{\sum_{i=t_1}^{t_k} v_i}{t_k - t_1 + 1}
\end{equation}

\begin{equation}\label{eq2:std}
    \begin{split}
    f_{std}(t_1,t_k) = \sqrt{\frac{\sum_{i=t_1}^{t_k} (v_i-f_{mean}(t_1,t_k))^2}{t_k - t_1}} \text{,~~~~~~~~~~~}
    \\
    \text{~~~where }t_2 > t_1
    \end{split}
\end{equation}

\begin{equation}\label{eq3:slope}
f_{slope}(t_1,t_k) = \widehat{\beta} 
\end{equation}
where $t_2 > t_1$ and $\widehat{\beta}$ is the slope of the least squares regression line in candidate interval sets.

The secondary transformation involves an extra localized pooling process applied to the interval features obtained from each individual time series. This operation gathers all the potential interval sets derived from the same time series and applies a statistical function (such as maximum, minimum, or mean) for aggregation. In other words, during this phase, we calculate the highest, lowest, and average values of statistical attributes (i.e., mean, standard deviation, and slope) from the subseries obtained through sliding window operations.

\begin{figure*}
    \centering
    \includegraphics[width=0.99\textwidth]{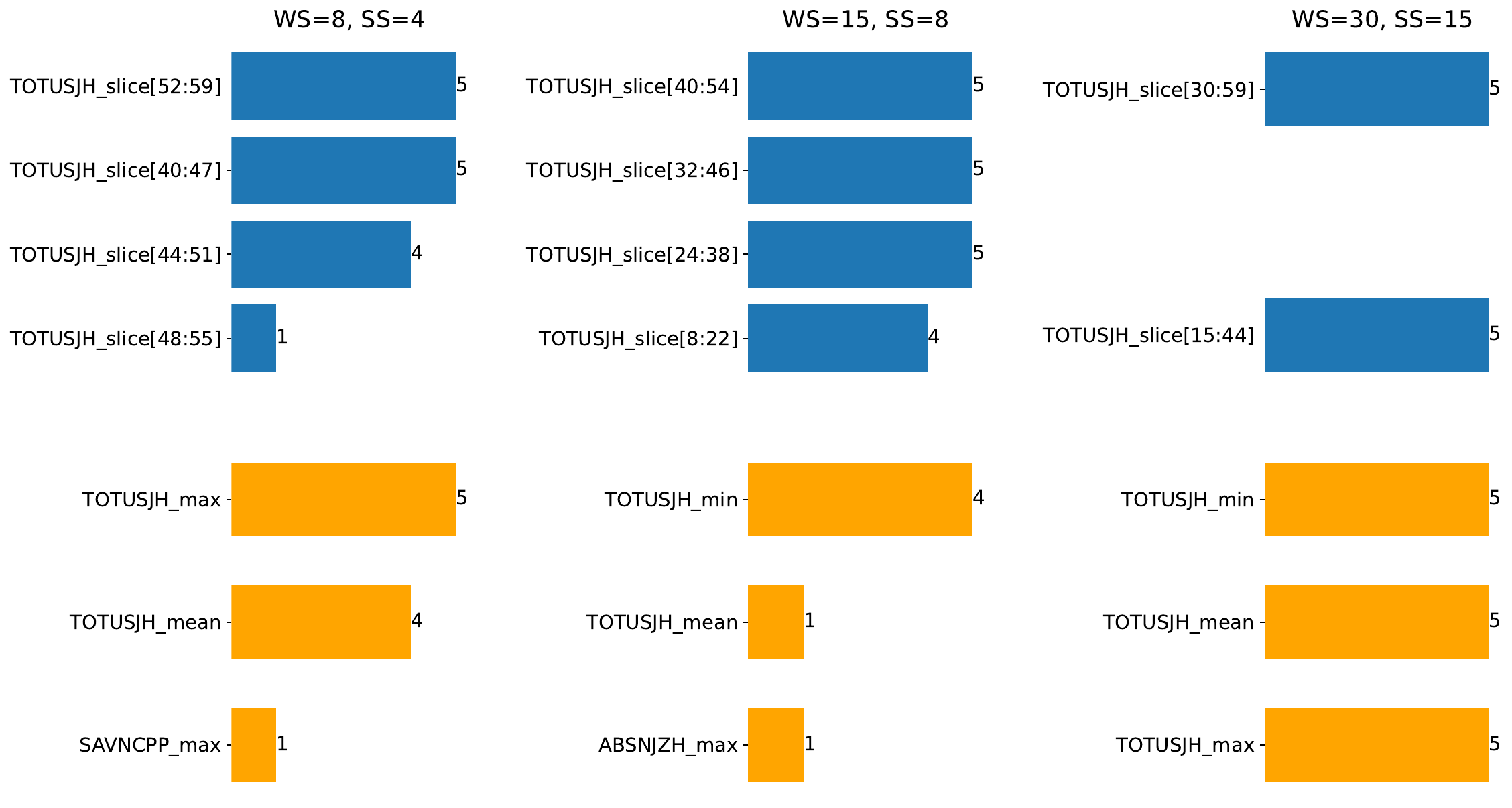}
    \caption{Results of our feature ranking experiments with only mean ($f_{mean}$ of slices used in the feature set and secondary transformation applied. Results come from an aggregation of five different imbalance weight settings and bars show how many times a feature has appeared in the top-5 ranking list. Orange bars show the features obtained after secondary transformations while blue ones show the local features. Note that WS represents the window size while SS represents the step size of candidate interval settings.}
    \label{fig:top5_mean}
\end{figure*}

\begin{figure*}[t]
\vspace{2em}
    \centering
    \includegraphics[width=0.99\textwidth]{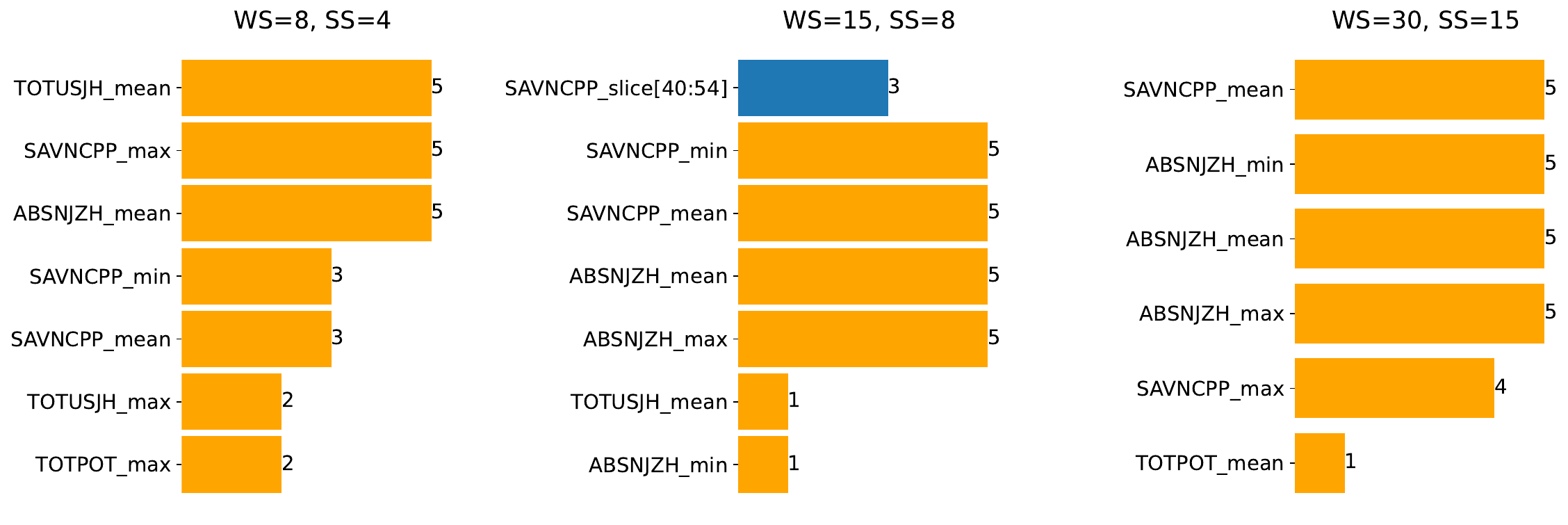}
     \caption{Results of our feature ranking experiments with only standard deviation ($f_{std}$ of slices used in the feature set and secondary transformation applied (similar to Fig.~\ref{fig:top5_mean}). Results come from an aggregation of five weight settings and bars show how many times a feature has appeared in the top-5 ranking list. Orange bars show the features obtained after secondary transformations and blue ones show the local features.}
     \label{fig:top5_std}
\vspace{2em}
     \centering
     \includegraphics[width=0.99\textwidth]{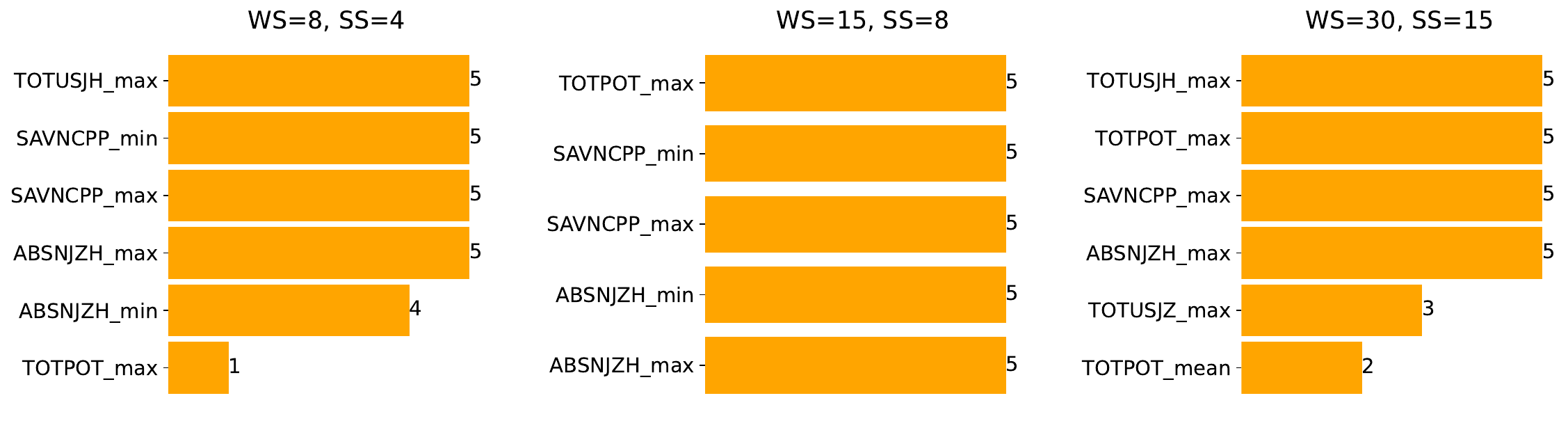}
     \caption{Results of our feature ranking experiments with only slope ($f_{slope}$ of slices used in the feature set and secondary transformation applied (similar to Fig.~\ref{fig:top5_mean}). Results come from an aggregation of five weight settings and bars show how many times a feature has appeared in the top-5 ranking list. Orange bars show the features obtained after secondary transformations and no local feature was ranked in the top 5 across all the experimental runs.} 
     \label{fig:top5_slope}
\end{figure*}

\subsection{Time Series Forest and Feature Ranking}
After deriving interval features from subsequences and applying a secondary transformation to these statistical features, we construct an input vector. This vector is then utilized for constructing a random forest, denoted as the \emph{sliding window time series forest}. While various classifiers could be employed for generating predictions, we opt for random forest classifiers for two primary reasons: (1) their effectiveness and robustness with noisy high-dimensional data, and (2) their inherent feature selection capabilities. It's worth noting that depending on the chosen parameter settings, such as employing smaller window and step sizes, the data space for the interval feature vectors can undergo significant expansion. Additionally, the process of vectorization based on the sliding window approach can produce data points that are relatively correlated and potentially noisy. Consequently, it becomes crucial to effectively eliminate these features through information-theoretic relevance metrics measures (e.g., entropy and Gini index, etc) to ensure the effectiveness of our approach.

Next, the feature importance scores obtained through the proposed sliding window time series forest are employed to establish a ranking for interval features. Different alternatives for the importance scores have been documented in the literature, such as the mean decrease in impurity or feature permutation. These scores enable us to rank the transformed features vector extracted from intervals within a framework of embedded feature selection scenarios. The classification based on random forests provides an enhancement of the robustness of classification, especially in tasks that need to be highly generalized.

\section{Experimental Evaluations}\label{experiments}
The experiments are structured with two objectives: (1) Showcasing the effectiveness and conducting a performance comparison among time series classifiers developed using distinct extracted interval features. (2) Identifying the intervals that are most relevant to the initial time series. Our goal is centered around offering interpretable insights into our model. This entails pinpointing the segments of the time series that influence predictions, along with comprehending the aggregation strategies that could yield more advantageous outcomes.

\subsection{Data Collection}
For our solar flare prediction study, we used an open source multivariate time series dataset named \textbf{S}pace \textbf{W}eather \textbf{AN}alytics dataset for \textbf{S}olar \textbf{F}lare prediction (SWAN-SF) by \cite{Angryk2020} as our benchmark dataset. This dataset provides space weather-related physical parameters derived from solar magnetograms onboarding the various observatory resources. We utilized six solar active region parameters which are: the total unsigned magnetic flux (USFLUX), the total unsigned vertical current (TOTUSJZ), the total unsigned current helicity (TOTUSJH), and the total photospheric magnetic free-energy density (TOTPOT), Sum of the Absolute Value of the Net Currents Per Polarity (SAVNCPP), Absolute value of the net current helicity (ABSNJZH). These are considered to be the most representative features of solar activity in \cite{Bobra2015}. Detailed descriptions of these parameters can be found on \cite{Bobra2015} \cite{Angryk2020}. For time and space consideration, we only use these six physical parameters, although the analysis can be extended with many more parameters.

The SWAN-SF dataset consists of five time-segmented partitions meaning that the data instances within each partition do not overlap with others in the time domain. The active regions are sliced through a sliding observation window (i.e., 12-hour intervals) over the multivariate time series. Each slice has a corresponding active region number (i.e., National Oceanic and Atmospheric Administration (NOAA) Active Region number matched with HMI Active Region Patches (HARPs) identifiers), a class label (i.e., the maximum intensity flare originating from that active region within the next 24 hours), a start and end timestamp of each slice. Note here that the flare intensity is determined by the logarithmic classification of peak X-ray flux. Within the major flaring categories (X, M, C, B, or A) from such schema, we include instances labeled with M- and X-class flares as the flaring (i.e., positive class) while the relatively weak C- and B- class flare labels and flare-quiet regions are considered as non-flaring (i.e., negative class). In this way, the flare forecasting problem can be modeled as a binary multivariate time series classification task.

Throughout the studies, we use Partitions 1,2,3, and 5 for training purposes and Partition 4 for testing purposes as it has a relatively more balanced flaring to non-flaring class ratio.

\subsection{Experimental Settings}
To evaluate the performance of our model, we employed a binary $2\times2$ contingency matrix along with various evaluation metrics and crucial forecast skill scores. Within these evaluation metrics, the positive class corresponds to the occurrence of a large flare class (i.e., $\ge$M1.0 in the NOAA/GOES flare classification) while the negative class refers to the relatively small flare and flare-quite region (i.e., $<$M1.0 class flare). Therefore, true positives (TPs) denote cases where the model accurately predicts a flare event (positive class), and true negatives (TNs) refer to instances where the model correctly predicts a small flare event (negative class). In contrast, false positives (FPs) represent false alarms, occurring when the model predicts a non-flaring event as flaring, and false negatives (FNs) indicate misses, which happen when the model fails to predict an actual flare event. In our study, we employ two well-known skill scores: the True Skill Statistic score (TSS) and the Heidke Skill Score (HSS). The former can be calculated as the disparity between the probability of detection (recall for the positive class) and the probability of false detection (POFD). This measurement is determined using the equation specified as Eq.~\ref{eq:tss}.

\begin{equation}
    \label{eq:tss}
        TSS = \frac{TP}{TP + FN} - \frac{FP}{FP + TN}
\end{equation}

The second measurement is the Heidke Skill Score and assesses how much the forecast enhances compared to a random forecast. This score varies from -$\infty$ to 1, where a score of 1 signifies flawless performance and 0 denotes a lack of skill. A score of 0 signifies that the forecast's accuracy is no better than a random binary forecast based on the provided class distributions. This metric can be computed using Eq.~\ref{eq:hss2}, where $P$ represents the sum of true positives ($TP$) and false negatives ($FN$), and $N$ corresponds to the sum of false positives ($FP$) and true negatives ($TN$) in the observed data.

\begin{equation}
    \label{eq:hss2}
        HSS = \frac{2 \cdot ((TP \cdot TN) - (FN \cdot FP))}{P \cdot (FN + TN) + N \cdot (TP + FP)}
\end{equation}

\subsection{Comparison between different window sizes}
To identify the potential candidate intervals, we employ a fixed-size sliding window approach with three configurations: window sizes of 8, 15, and 30, and corresponding step sizes of 4, 8, and 15, respectively. This approach extracts intervals from the original time series based on pre-defined window sizes (WS) and traverses the time series using the designated step size (SS). These identified intervals are then transformed with statistical functions (i.e., mean, standard deviation, slope) to generate informative interval features. To assess the predictive capabilities of these extracted features, we train a Random Forest (RF) classifier for each window setting. The outcomes are presented in Table \ref{table:eval_ws_ss}, showcasing the forecast skill scores in terms of TSS, HSS and recall.

\begin{table}[htbp]

\caption{The evaluation metrics for Random Forest with different window sizes and step sizes settings}  
\renewcommand{\arraystretch}{2.25}
    \setlength{\tabcolsep}{4pt}
    \centering
    \begin{tabular}{r c c c}
    \hline
    Model &
    TSS &
    HSS &
    Recall 
     \\ \hline
    \begin{minipage}{0.13\textwidth} \raggedleft WS=60 \end{minipage} &
      0.16 &
      0.27 &
      0.16 \\ 
    \begin{minipage}{0.13\textwidth} \raggedleft WS=8, SS=4 \end{minipage} &
      0.82 &
      0.19 &
      0.97 \\ 
    \begin{minipage}{0.13\textwidth} \raggedleft 
    WS=15, SS=8 \end{minipage} &
      0.83 &
      0.21 &
      0.97 \\ 
    \begin{minipage}{0.13\textwidth} \raggedleft 
    WS=30, SS=15 \end{minipage} &
      0.83 &
      0.20 &
      0.98 \\ \hline
    \end{tabular}
    \label{table:eval_ws_ss}
\end{table}

Results in Table \ref{table:eval_ws_ss} show that the random forest models constructed using different configurations exhibit consistent performance across different window sizes in terms of TSS and HSS. These models achieve an average TSS of 82.3\% and an average HSS of 20\%. Notably, these models are trained on a substantially imbalanced class weight, with an imbalance ratio of approximately 1:50 between our positive (flaring) and negative (non-flaring) classes, based on the original SWAN-SF benchmark dataset. Given this situation, the subsequent experiment involves adjusting the class weights to achieve a balanced class distribution and to identify pertinent interval features.

\subsection{Class weights tuning}
In this part of the study, we conduct an experiment involving the training of five Random Forest (RF) models. These models vary in terms of the combinations of class weights assigned to our flaring and non-flaring classes. The primary aim is to reveal interval features that hold greater significance for a predictive task, thereby facilitating the interpretation of the model's feature selection process. The models are all trained using consistent sliding window settings, along with the separate utilization of each statistical function. The outcomes obtained from all the combinations are aggregated to generate a ranking list based on the occurrence frequency of the top 5 features. The visual representation of these results is provided in Figures \ref{fig:top5_mean} through \ref{fig:top5_slope}.

As depicted in Figure \ref{fig:top5_mean}, the blue bars represent the identified relevant intervals whose features contribute the most to the prediction, while the orange bars signify the relevant intervals and features from the secondary transformation. The observation from Figure \ref{fig:top5_mean} highlights that the segments starting at the index of 40 within the TOTUSJH (the total unsigned current helicity) parameter are frequently repeated across all window settings, which suggests a potential starting point for the most relevant intervals. Considering we use a 12-min cadence and length-60 time series, this data point corresponds to approximately 4 hours before the end time of the observation window.

Upon employing secondary transformations, we observe that distinct patterns emerge. For instance, we note a strong repetitive pattern of occurrences of the minimum/mean values of TOTUSJH within a size-8 window and size-4 steps (in terms of timestamps), the minimum value of SAVNCPP (Sum of the Absolute Value of the Net Currents Per Polarity) within a size-15 window and size-8 step, as well as the maximum value of TOTUSJH within a size-30 window and size-15 step. This occurrence frequency implies that these physical parameters could serve as promising candidates for temporal dynamics leading to a relatively large flare event.

We also observe that only a minimal number of local slice features from standard deviation and slope features are considered significant for solar flare prediction. This suggests that our classifiers' predictions do not exclusively rely on the interval features from individual slices but also consider features we obtain after secondary transformation.

\section{Conclusions and Future Work}\label{conclusions}
In this study, we have undertaken the training of classifiers for multivariate time series using intervals and introduced an innovative approach for ranking features obtained from sliding window intervals. Our work concentrates on including interpretability in interval-based multivariate time series classifiers. Our work's primary achievement is in connecting the black-box models used for high-dimensional data and investigating the key sub-intervals within multivariate time series, particularly their relevance in the context of solar flare prediction. Our findings demonstrate a skill score of considerable effectiveness across various window size configurations. Furthermore, we present an outline of our ranking framework, highlighting the significance of identified statistical features in predicting solar flares.

There are a few potential directions available for future research. These include the extension of the sliding time window dimensions and step sizes, the incorporation of alternative ranking metrics and physical parameters, as well as the exploration of diverse time series models.

\section*{Acknowledgment}
 This work is supported in part under two grants from NSF (Award \#2104004) and NASA (SWR2O2R Grant \#80NSSC22K0272).

\bibliographystyle{splncs04}
\bibliography{mybib}

\begin{thebibliography}{10}
\providecommand{\url}[1]{\texttt{#1}}
\providecommand{\urlprefix}{URL }
\providecommand{\doi}[1]{https://doi.org/#1}

\bibitem{Angryk2020}
Angryk, R.A., Martens, P.C., Aydin, B., Kempton, D., Mahajan, S.S., Basodi, S.,
  Ahmadzadeh, A., Cai, X., Boubrahimi, S.F., Hamdi, S.M., Schuh, M.A.,
  Georgoulis, M.K.: Multivariate time series dataset for space weather data
  analytics. Scientific Data  \textbf{7}(1) (Jul 2020).
  \doi{10.1038/s41597-020-0548-x},
  \url{https://doi.org/10.1038/s41597-020-0548-x}

\bibitem{Bagnall2016}
Bagnall, A., Lines, J., Bostrom, A., Large, J., Keogh, E.: The great time
  series classification bake off: a review and experimental evaluation of
  recent algorithmic advances. Data Mining and Knowledge Discovery
  \textbf{31}(3),  606--660 (Nov 2016). \doi{10.1007/s10618-016-0483-9},
  \url{https://doi.org/10.1007/s10618-016-0483-9}

\bibitem{6497440}
Baydogan, M.G., Runger, G., Tuv, E.: A bag-of-features framework to classify
  time series. IEEE Transactions on Pattern Analysis and Machine Intelligence
  \textbf{35}(11),  2796--2802 (2013). \doi{10.1109/TPAMI.2013.72}

\bibitem{Benz2008}
Benz, A.O.: Flare observations. Living Reviews in Solar Physics  \textbf{5}
  (2008). \doi{10.12942/lrsp-2008-1},
  \url{https://doi.org/10.12942/lrsp-2008-1}

\bibitem{Bobra2015}
Bobra, M.G., Couvidat, S.: Solar flare prediction using sdo/hmi vector magnetic
  field data with a machine-learning algorithm. The Astrophysical Journal
  \textbf{798}(2), ~135 (Jan 2015). \doi{10.1088/0004-637x/798/2/135},
  \url{https://doi.org/10.1088/0004-637x/798/2/135}

\bibitem{Bobra2016}
Bobra, M.G., Ilonidis, S.: {PREDICTING} {CORONAL} {MASS} {EJECTIONS} {USING}
  {MACHINE} {LEARNING} {METHODS}. The Astrophysical Journal  \textbf{821}(2),
  ~127 (Apr 2016). \doi{10.3847/0004-637x/821/2/127},
  \url{https://doi.org/10.3847/0004-637x/821/2/127}

\bibitem{Boucheron2015}
Boucheron, L.E., Al-Ghraibah, A., McAteer, R.T.J.: Prediction of solar flare
  size and time-to-flare using support vector machine regression. The
  Astrophysical Journal  \textbf{812}(1), ~51 (Oct 2015).
  \doi{10.1088/0004-637x/812/1/51},
  \url{https://doi.org/10.1088/0004-637x/812/1/51}

\bibitem{9378006}
Chen, Y., Ji, A., Babajiyavar, P.A., Ahmadzadeh, A., Angryk, R.A.: On the
  effectiveness of imaging of time series for flare forecasting problem. In:
  2020 IEEE International Conference on Big Data (Big Data). pp. 4184--4191
  (2020). \doi{10.1109/BigData50022.2020.9378006}

\bibitem{Deng2013}
Deng, H., Runger, G., Tuv, E., Vladimir, M.: A time series forest for
  classification and feature extraction. Information Sciences  \textbf{239},
  142--153 (Aug 2013). \doi{10.1016/j.ins.2013.02.030},
  \url{https://doi.org/10.1016/j.ins.2013.02.030}

\bibitem{DBLP:conf/pkdd/EruhimovMT07}
Eruhimov, V., Martyanov, V., Tuv, E.: Constructing high dimensional feature
  space for time series classification. In: Kok, J.N., Koronacki, J.,
  de~M{\'{a}}ntaras, R.L., Matwin, S., Mladenic, D., Skowron, A. (eds.)
  Knowledge Discovery in Databases: {PKDD} 2007, 11th European Conference on
  Principles and Practice of Knowledge Discovery in Databases, Warsaw, Poland,
  September 17-21, 2007, Proceedings. Lecture Notes in Computer Science,
  vol.~4702, pp. 414--421. Springer (2007).
  \doi{10.1007/978-3-540-74976-9\_41},
  \url{https://doi.org/10.1007/978-3-540-74976-9\_41}

\bibitem{Fletcher2011}
Fletcher, L., Dennis, B.R., Hudson, H.S., Krucker, S., Phillips, K., Veronig,
  A., Battaglia, M., Bone, L., Caspi, A., Chen, Q., Gallagher, P., Grigis,
  P.T., Ji, H., Liu, W., Milligan, R.O., Temmer, M.: An observational overview
  of solar flares. Space Science Reviews  \textbf{159}(1-4),  19--106 (Aug
  2011). \doi{10.1007/s11214-010-9701-8},
  \url{https://doi.org/10.1007/s11214-010-9701-8}

\bibitem{Georgoulis2012}
Georgoulis, M.K.: On our ability to predict major solar flares. In: The Sun:
  New Challenges, pp. 93--104. Springer Berlin Heidelberg (2012).
  \doi{10.1007/978-3-642-29417-4\_9},
  \url{https://doi.org/10.1007/978-3-642-29417-4\_9}

\bibitem{Geurts2001}
Geurts, P.: Pattern extraction for time series classification. In: Principles
  of Data Mining and Knowledge Discovery, pp. 115--127. Springer Berlin
  Heidelberg (2001). \doi{10.1007/3-540-44794-6\_10},
  \url{https://doi.org/10.1007/3-540-44794-6\_10}

\bibitem{Homayouni2020}
Homayouni, H., Ghosh, S., Ray, I., Gondalia, S., Duggan, J., Kahn, M.G.: An
  autocorrelation-based {LSTM}-autoencoder for anomaly detection on time-series
  data. In: 2020 {IEEE} International Conference on Big Data (Big Data). {IEEE}
  (Dec 2020). \doi{10.1109/bigdata50022.2020.9378192},
  \url{https://doi.org/10.1109/bigdata50022.2020.9378192}

\bibitem{Hong2023}
Hong, J., Ji, A., Pandey, C., Aydin, B.: Beyond traditional flare forecasting:
  A data-driven labeling approach for~high-fidelity predictions. In: Big Data
  Analytics and Knowledge Discovery, pp. 380--385. Springer Nature Switzerland
  (2023). \doi{10.1007/978-3-031-39831-5\_34},
  \url{https://doi.org/10.1007/978-3-031-39831-5_34}

\bibitem{Huang2010}
Huang, X., Yu, D., Hu, Q., Wang, H., Cui, Y.: Short-term solar flare prediction
  using predictor teams. Solar Physics  \textbf{263}(1-2),  175--184 (Apr
  2010). \doi{10.1007/s11207-010-9542-3},
  \url{https://doi.org/10.1007/s11207-010-9542-3}

\bibitem{9750381}
Ji, A., Arya, A., Kempton, D., Angryk, R., Georgoulis, M.K., Aydin, B.: A
  modular approach to building solar energetic particle event forecasting
  systems. In: 2021 IEEE Third International Conference on Cognitive Machine
  Intelligence (CogMI). pp. 106--115 (2021).
  \doi{10.1109/CogMI52975.2021.00022}

\bibitem{9377906}
Ji, A., Aydin, B., Georgoulis, M.K., Angryk, R.: All-clear flare prediction
  using interval-based time series classifiers. In: 2020 IEEE International
  Conference on Big Data (Big Data). pp. 4218--4225 (2020).
  \doi{10.1109/BigData50022.2020.9377906}

\bibitem{Karlsson2016}
Karlsson, I., Papapetrou, P., Bostr\"{o}m, H.: Generalized random shapelet
  forests. Data Mining and Knowledge Discovery  \textbf{30}(5),  1053--1085
  (Jul 2016). \doi{10.1007/s10618-016-0473-y},
  \url{https://doi.org/10.1007/s10618-016-0473-y}

\bibitem{Kusano2020}
Kusano, K., Iju, T., Bamba, Y., Inoue, S.: A physics-based method that can
  predict imminent large solar flares. Science  \textbf{369}(6503),  587--591
  (Jul 2020). \doi{10.1126/science.aaz2511},
  \url{https://doi.org/10.1126/science.aaz2511}

\bibitem{Lines2014}
Lines, J., Bagnall, A.: Time series classification with ensembles of elastic
  distance measures. Data Mining and Knowledge Discovery  \textbf{29}(3),
  565--592 (Jun 2014). \doi{10.1007/s10618-014-0361-2},
  \url{https://doi.org/10.1007/s10618-014-0361-2}

\bibitem{Lubba2019}
Lubba, C.H., Sethi, S.S., Knaute, P., Schultz, S.R., Fulcher, B.D., Jones,
  N.S.: catch22: {CAnonical} time-series {CHaracteristics}. Data Mining and
  Knowledge Discovery  \textbf{33}(6),  1821--1852 (Aug 2019).
  \doi{10.1007/s10618-019-00647-x},
  \url{https://doi.org/10.1007/s10618-019-00647-x}

\bibitem{DBLP:journals/corr/abs-2008-09172}
Middlehurst, M., Large, J., Bagnall, A.J.: The canonical interval forest
  {(CIF)} classifier for time series classification. CoRR
  \textbf{abs/2008.09172} (2020), \url{https://arxiv.org/abs/2008.09172}

\bibitem{10.5555/766914.766918}
Nanopoulos, A., Alcock, R., Manolopoulos, Y.: Feature-Based Classification of
  Time-Series Data, p. 49–61. Nova Science Publishers, Inc., USA (2001)

\bibitem{9671322}
Pandey, C., Angryk, R.A., Aydin, B.: Solar flare forecasting with deep neural
  networks using compressed full-disk hmi magnetograms. In: 2021 IEEE
  International Conference on Big Data (Big Data). pp. 1725--1730 (2021).
  \doi{10.1109/BigData52589.2021.9671322}

\bibitem{pandey2023interpretable}
Pandey, C., Ji, A., Angryk, R.A., Aydin, B.: Towards interpretable solar flare
  prediction with attention-based deep neural networks (2023)

\bibitem{Pandey2022}
Pandey, C., Ji, A., Angryk, R.A., Georgoulis, M.K., Aydin, B.: Towards coupling
  full-disk and active region-based flare prediction for operational space
  weather forecasting. Frontiers in Astronomy and Space Sciences  \textbf{9}
  (Aug 2022). \doi{10.3389/fspas.2022.897301},
  \url{https://doi.org/10.3389/fspas.2022.897301}

\bibitem{priest2002magnetic}
Priest, E., Forbes, T.: The magnetic nature of solar flares. The Astronomy and
  Astrophysics Review  \textbf{10}(4),  313--377 (2002)

\bibitem{1163055}
Sakoe, H., Chiba, S.: Dynamic programming algorithm optimization for spoken
  word recognition. IEEE Transactions on Acoustics, Speech, and Signal
  Processing  \textbf{26}(1),  43--49 (1978). \doi{10.1109/TASSP.1978.1163055}

\bibitem{shibata2011solar}
Shibata, K., Magara, T.: Solar flares: magnetohydrodynamic processes. Living
  Reviews in Solar Physics  \textbf{8}(1), ~6 (2011)

\bibitem{Silva2018}
Silva, D.F., Giusti, R., Keogh, E., Batista, G.E.A.P.A.: Speeding up similarity
  search under dynamic time warping by pruning unpromising alignments. Data
  Mining and Knowledge Discovery  \textbf{32}(4),  988--1016 (Mar 2018).
  \doi{10.1007/s10618-018-0557-y},
  \url{https://doi.org/10.1007/s10618-018-0557-y}

\bibitem{Song2008}
Song, H., Tan, C., Jing, J., Wang, H., Yurchyshyn, V., Abramenko, V.:
  Statistical assessment of photospheric magnetic features in imminent solar
  flare predictions. Solar Physics  \textbf{254}(1),  101--125 (Nov 2008).
  \doi{10.1007/s11207-008-9288-3},
  \url{https://doi.org/10.1007/s11207-008-9288-3}

\bibitem{Ye2010}
Ye, L., Keogh, E.: Time series shapelets: a novel technique that allows
  accurate, interpretable and fast classification. Data Mining and Knowledge
  Discovery  \textbf{22}(1-2),  149--182 (Jun 2010).
  \doi{10.1007/s10618-010-0179-5},
  \url{https://doi.org/10.1007/s10618-010-0179-5}

\bibitem{Yu2009}
Yu, D., Huang, X., Wang, H., Cui, Y.: Short-term solar flare prediction using a
  sequential supervised learning method. Solar Physics  \textbf{255}(1),
  91--105 (Feb 2009). \doi{10.1007/s11207-009-9318-9},
  \url{https://doi.org/10.1007/s11207-009-9318-9}

\end{thebibliography}

\end{document}